\documentclass[amsmath,amssymb,aps,prb,twocolumn,superscriptaddress,10pt,floatfix]{revtex4-2}

\usepackage{graphicx}
\usepackage{bm}
\usepackage[normalem]{ulem}

\begin{document}

\title{Magnetic self-frustration from spontaneous structural distortion}

\author{T. Vignau Costa}
\affiliation{Instituto de Física de Líquidos y Sistemas Biológicos (IFLYSIB), UNLP-CONICET, Argentina.} 
\affiliation{Departamento de Física, Facultad de Ciencias Exactas, Universidad Nacional de La Plata, Argentina.}

\author{G. L. Rossini}
\affiliation{Instituto de Física de Líquidos y Sistemas Biológicos (IFLYSIB), UNLP-CONICET, Argentina.} 
\affiliation{Departamento de Física, Facultad de Ciencias Exactas, Universidad Nacional de La Plata, Argentina.}

\author{D. C. Cabra}
\affiliation{Instituto de Física de Líquidos y Sistemas Biológicos (IFLYSIB), UNLP-CONICET, Argentina.} 
\affiliation{Departamento de Física, Facultad de Ciencias Exactas, Universidad Nacional de La Plata, Argentina.}
\author{S. A. Grigera}
\affiliation{Instituto de Física de Líquidos y Sistemas Biológicos (IFLYSIB), UNLP-CONICET, Argentina.} 
\affiliation{Departamento de Física, Facultad de Ciencias Exactas, Universidad Nacional de La Plata, Argentina.}
\author{R. A. Borzi}
\affiliation{Instituto de Física de Líquidos y Sistemas Biológicos (IFLYSIB), UNLP-CONICET, Argentina.} 
\affiliation{Departamento de Física, Facultad de Ciencias Exactas, Universidad Nacional de La Plata, Argentina.}

\date{\today}

\begin{abstract}

In frustrated magnetism, lattice distortions mediated by magnetoelastic coupling are commonly invoked as an escape route from extensive degeneracy toward an ordered ground state and, in some cases, the onset of multiferroicity.  Here we present a minimal classical model that illustrates  the converse phenomenon, 
which we term ``magnetic self-frustration''. Monte Carlo simulations reveal that a kagom\'e lattice with trivial magnetic interactions---namely, nearest-neighbor Ising ferromagnetism---undergoes a magnetostructural transition into a breathing-like phase, characterized by irregular bond dimerization along the three kagom\'e directions. Structurally, the equilateral triangles belonging to one of the two kagom\'e sublattices spontaneously distort, expanding into isosceles triangles. An analysis to first order in the magnetoelastic coupling constant $\alpha$ shows that the shape of these triangles is remarkably robust. Acting like rigid building blocks in a puzzle, their vertices determine the geometry of the second sublattice, giving rise to contracted ferromagnetic triangles with a variety of shapes. The magnetic sector can be mapped onto an effective antiferromagnetic triangular lattice, which remains disordered down to low temperatures and retains a finite residual entropy of one third of that of Wannier.  This self-frustrated phase takes place at intermediate values of $\alpha$, separating the conventional undistorted ferromagnetic phase at weak coupling from a strongly coupled ordered phase characterized by a regular dimerized up--up--down--down antiferromagnetic pattern along the three kagomé directions, built from ferromagnetic triangles and hexagons.
\end{abstract}
	
\maketitle

\section{Introduction}

Magnetically frustrated materials are known to exhibit a wide range of unconventional physical phenomena~\cite{lacroix2011introduction,diep2013frustrated}, including finite residual entropy at zero temperature~\cite{moessner1998low}, scale-free correlations in the absence of a conventional phase transition, exotic excitations~\cite{balents2010spin}, topological transitions~\cite{jaubert2008three,pili2022topological,szabo2025hidden}, order-by-disorder mechanisms~\cite{lubensky2011order,moessner1998low,guruciaga2016field}, and multiferroicity~\cite{mostovoy2007multiferroics}. 
Such behavior originates from strong competition between pairwise interactions, which can arise through several distinct mechanisms, beyond the quenched disorder characteristic of spin glasses~\cite{mydosh2015spin}.
One route is energetic: for a fixed lattice, exchange interactions extending over different length scales (first, second, third nearest neighbors, etc.) can be tuned to produce a highly degenerate ground-state manifold. 
A second, and often more robust, mechanism is geometric: even if we limit to nearest neighbor interactions, the lattice geometry itself can enforce frustration and extensive degeneracy over a broad range of interaction parameters, a situation known as \textit{geometrical frustration} \cite{ramirez2001geometrical,moessner2006geometrical}. 
In this work, we explore a third, less conventional route to frustration. In contrast to the conventional paradigm, where lattice distortions relieve frustration (e.g. \cite{ji2009spin,frandsen2020nanoscale}), we show that magnetoelastic coupling can lead to frustration, even in an otherwise non frustrated system.

The kagom\'e lattice is one of the most extensively studied two-dimensional geometries. It consists of a network of corner-sharing equilateral triangles, a structure that naturally gives rise to strong geometric frustration and highly nontrivial collective behavior~\cite{yin2022topological,teng2022discovery,negi2025magnetic,broholm2020quantum,di2026kagome,wang2023quantum}.   Kagom\'e planes also appear as building blocks of several three-dimensional materials, most notably pyrochlore lattices, where similar frustration effects play an important role~\cite{Borzi2016}. Depending on the microscopic degrees of freedom and the nature of their interactions, kagom\'e systems can host a broad range of exotic classical and quantum phases. The large degeneracy associated with frustration also makes these systems especially sensitive to perturbations such as anisotropy, disorder, and coupling to the lattice.

Here we consider Ising spins located on the sites of a kagom\'e lattice coupled to local elastic distortions. Small lattice deformations modify the distances between neighboring spins and therefore alter the magnetic exchange interactions. The interplay between frustration, spin ordering, and lattice elasticity provides a natural mechanism for the emergence of complex magnetic behavior and symmetry breaking.

\begin{figure}[bt]
\centering
\includegraphics[width=0.9\linewidth]{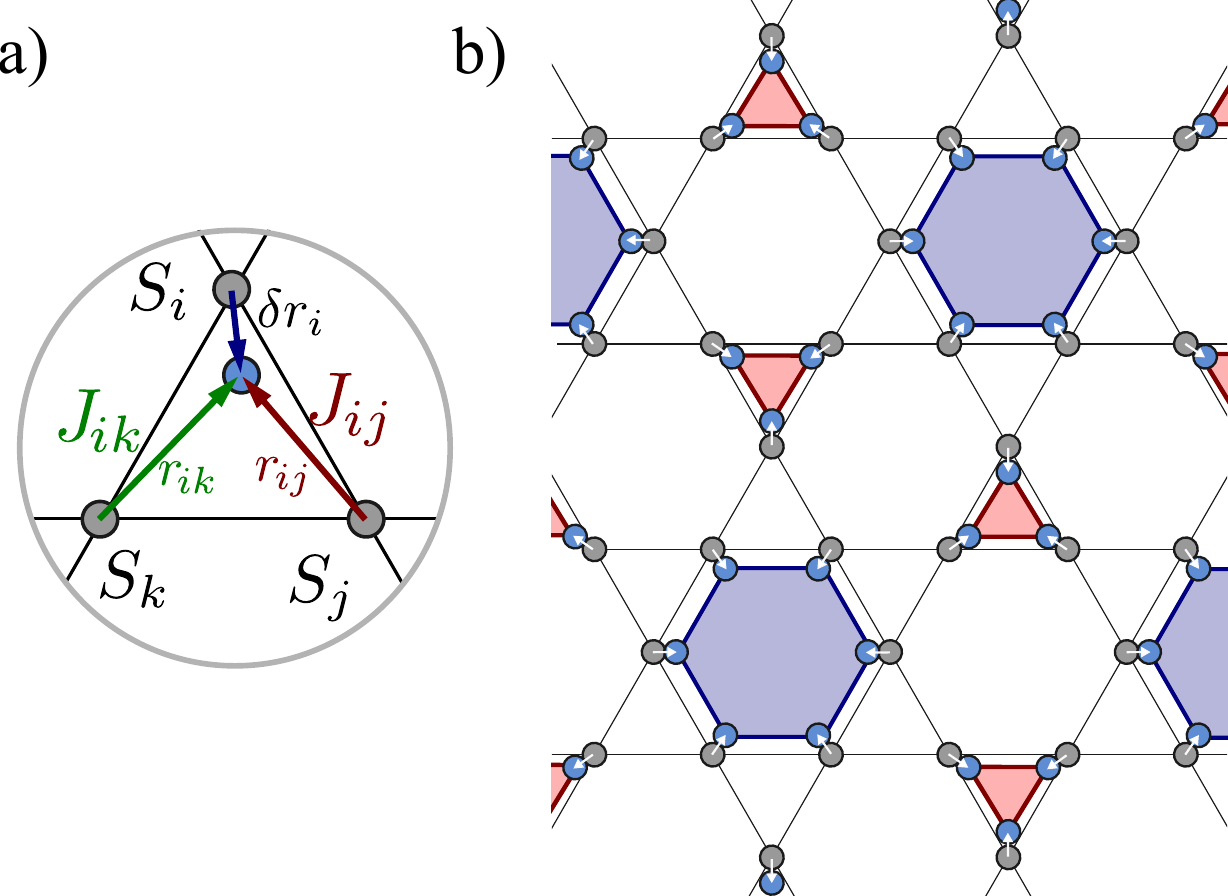}
\caption{ 
\textit{a-} 
A fragment of the kagomé lattice.
A distortion at site $i$ is indicated together with its effect on the neighboring exchange couplings
($J_{ij}=1-\alpha\, (r_{ij}-1)$, see Eq.~(\ref{eq:Jij})).
Bond-shortening favors ferromagnetism, while enough lengthening leads to antiferromagnetism. 
\textit{b-} 
The checkerboard configuration on the kagomé lattice, stable at large $\alpha$ values. 
Arrows indicate ion displacements from the pristine lattice (gray dots). 
The distorted configuration (blue dots) consists of contracted triangles and hexagons; 
along the three kagomé directions there is a length dimerization, seen as an alternation of short (ferro) and long (antiferro) bonds. 
Spins are aligned ferromagnetically in each contracted plaquette,  
where the orientation at triangles (in light red) is opposite to that at hexagons (in light blue);
in consequence, an up-up-down-down magnetic order is formed along the three kagomé directions. 
There is no frustration, as all the magnetic couplings are satisfied. 
}
\label{fig:structures}
\end{figure}

We have previously studied the effect of elastic distortions on a kagom\'e ice model, 
with Ising spins pointing in/out the lattice triangles~\cite{2013-kagome_ice}, as well as the ground states of a magnetoelastic Ising model on the square and kagom\'e lattices~\cite{pili2019two}.  On the ferromagnetic square lattice, contracted ferromagnetically aligned plaquettes were found above a critical magnetoelastic constant value,  forming a regular pattern dubbed checkerboard (CB). The lattice distortions lead to bond dimerization into ferro- and antiferro- magnetic exchange couplings along both axes.
On the kagomé lattice  contracted triangular and hexagonal ferromagnetic plaquettes were found (also dubbed checkerboard in Ref.~\cite{pili2019two}), see Fig.~\ref{fig:structures} here); similar spatially dimerized and  up-up-down-down magnetic orders are formed, now along the \emph{three} lattice symmetry axes. 
In both cases, no frustration emerges.

As we will discuss in this article,  a careful revision of the ferromagnetic kagom\'e geometry employing improved numerical techniques has revealed the existence of a third phase, intermediate between the uniform ferromagnetic and the regularly distorted CB phase.
Surprisingly, this phase is only partially ordered revealing the emergence of magnetic frustration, in spite of the trivial magnetic order associated with the original lattice. Due to this, we refer to it as the self-frustrated (SF) phase.  Notably, the self-frustration described here does not emerge from a contrived or finely tuned model, but rather from a system that can be regarded as a textbook example.
This is the main finding of this work, in which we also try to elucidate its origin and characterize its properties. 

The paper is organized as follows: 
In Section II we present the magneto-elastic model and Monte Carlo computation methods.
In Section III we present the Monte Carlo simulation results (including specific heat, entropy, lattice distortions and spin patterns, static structure factors).
Section IV describes at  microscopic level the quasi-degenerate ground state of the self-frustrated phase, and discusses its properties. 
A linear expansion of the interaction Hamiltonian provides an explanation of the energy degeneracy, while second order corrections set its limits. 
Finally, in Section V  we summarize our results and discuss further interpretations, relations and properties of the SF phase, from which we draw conclusions and perspectives.

\section{Model and Methods}

We consider a kagom\'e lattice, where magnetic ions carrying Ising spins can elastically displace 
from their regular positions. 
Such displacements are regulated by independent harmonic oscillators (the {\em Einstein site phonon spin model} \cite{wang2008app,gen2022nematicity}). 
Besides its simplicity, this is an appropriate model to describe optical phonons, 
expected to make a large contribution to lattice distortion when the active magnetic lattice is a part of a three-dimensional structure. 
The system is defined by the dimensionless Hamiltonian:
\begin{equation}
\frac{\mathcal{H}}{|J_0|}=\sum_{\langle i,j \rangle} J_{ij}(\textbf{r}_{ij}) S_i S_j  +\frac{K_e}{2} \sum_i (\delta r_{i})^2 
\label{eq:H}
\end{equation}
where $S_i=\pm 1$ represents an Ising pseudospin at site $i$ and $\langle i,j \rangle$ indicates nearest neighbor site pairs. 
The spin positions are given by $\textbf{r}_i$, measured in units such that the cell constant is $a=1$, 
while $\textbf{r}_{ij}=\textbf{r}_i-\textbf{r}_j$ (see Fig.~\ref{fig:structures}a)) are the relative position vectors of different spins. 
We call $\delta \textbf{r}_i \equiv \textbf{r}_i-\textbf{r}_i^0$ in the second term of Eq.~(\ref{eq:H}) the site displacement, while
$J_0$ is the exchange energy of the undistorted lattice, 
and $K_e$ is a stiff dimensionless elastic constant. 

We assume that the exchange coupling $J_{ij}$ (measured also in units of $J_0$) can be expanded to first order as a function of ions' distance change:
\begin{equation}
J_{ij}(\textbf{r}_{ij}) = \mathrm{sgn}(J_0)\,(1-\alpha \left( r_{ij}-1\right) ) \ ,
\label{eq:Jij}
\end{equation}
where $\alpha > 0$ is the magnetoelastic coupling and $r_{ij}=|\textbf{r}_{ij}|$ is the modified distance between ions $i$ and $j$.

Here we focus on the (undistorted) ferromagnetic (FM) case, assuming that $\mathrm{sgn}(J_0) = -1$.
A stretched bond ($r_{ij}>1$)  therefore weakens ferromagnetism and tends to reinforce antiferromagnetism 
(or, equivalently, antiferromagnetic alignment tends to push the spins apart).
Although this effect is written to depend solely on the interatomic distance, 
it can also be interpreted as arising from the angular dependence of a superexchange interaction. 

The elastic constant is set at $K_e=7200$; for this value the melting temperature of the crystal (according to Lindemans's criterion) 
and the Curie temperature (scaling with the exchange energy of the undistorted lattice $J_0$) will be separated by more than one order of magnitude~\cite{pili2019two}.

To parameterize the elastic distortions  $\delta \textbf{r}_i$ (see Fig.~\ref{fig:structures}) we use polar coordinates $(\rho_i,\theta_i)$.
The angular variable $\theta_i$ is treated as a clock model with $360$ equally spaced orientations, while the radial displacement $\rho_i$ is randomly chosen from an interval $[0,\delta_{\max}(T)]$ where the maximum amplitude $\delta_{\max}(T)$ depends on temperature. In accordance with the spirit of the model, the magnetic and elastic degrees of freedom are treated on equal standing in the Monte Carlo simulations~\cite{pili2019two}. Each Monte Carlo step is divided into an elastic update followed by a sequence of magnetic updates, allowing the magnetic degrees of freedom to equilibrate after each lattice move. Once full equilibration is reached, the results become insensitive to the specific ratio of magnetic to lattice updates.

In order to shorten equilibration times we have chosen to use a Parallel Tempering algorithm~\cite{newman1999monte}. 
To optimize the number of parallel configurations to be simulated and their temperature spacing, 
we performed a first simulation with 100 copies at equally spaced temperatures $T/|J_0|$ between $10^{-2}$ and $4$. 
We then used the spring algorithm~\cite{2012JCoPh.231.1524G} to determine a new set of temperatures, 
spacing them to guarantee a constant exchange probability of 0.6 between configurations at consecutive values of $T$.
In order to favor the independence between configurations we attempt swaps every 100 Monte Carlo steps. 
Larger temperatures were also simulated to characterize the fully disordered phase.
The results presented here were obtained for lattices with linear size $L=6$ kagom\'e unit cells. 
These simulations, corresponding to a total of $N = 6L^2 = 264$ spins,   
are already  computationally demanding, as a single step entails both elastic and magnetic moves. 


\section{Results}

\begin{figure}[bt]
\centering
\includegraphics[width=1.0\linewidth]{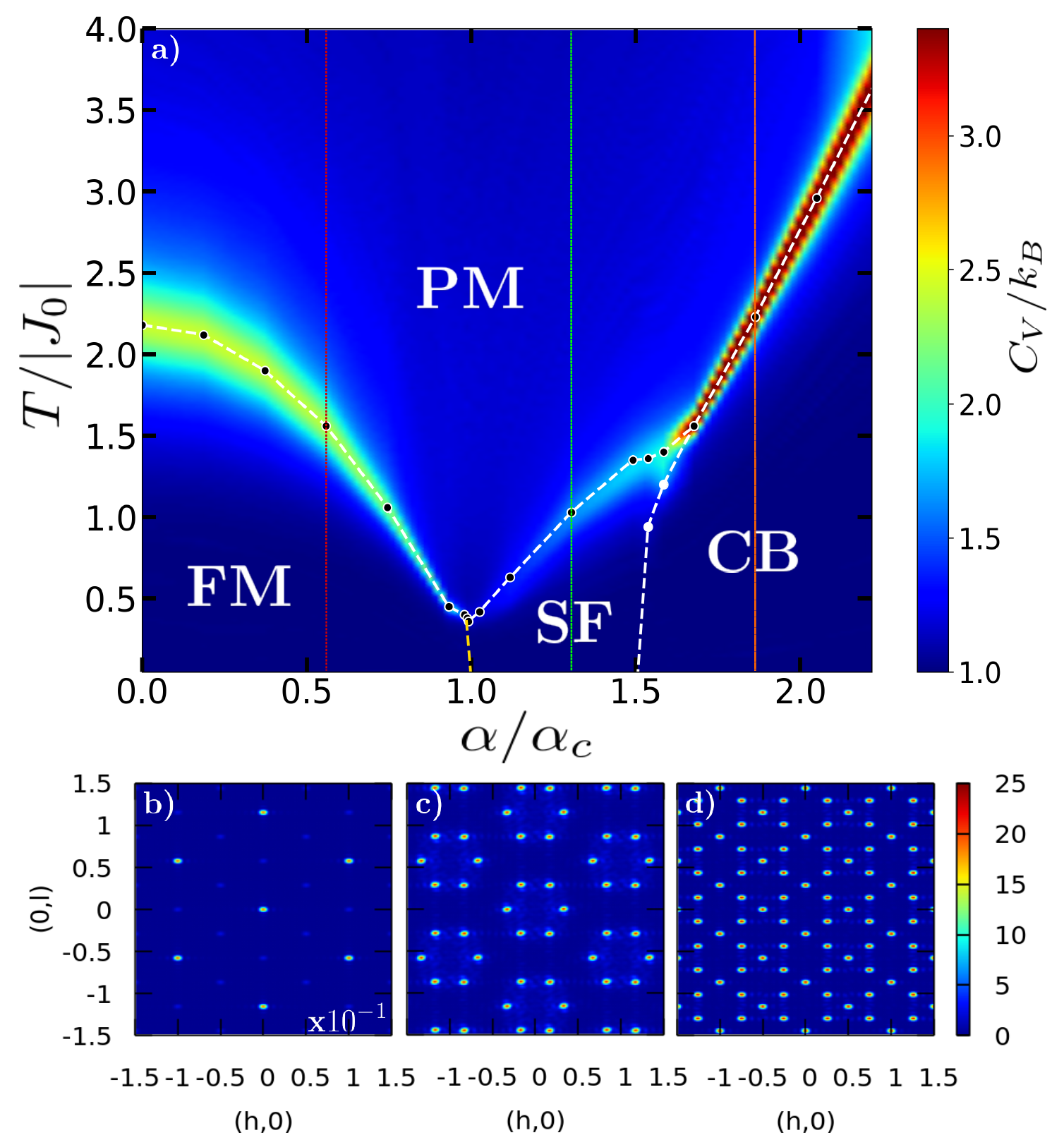}
\caption{
a) The phase diagram $\alpha - T$ for the elastic kagom\'e lattice. The color plot is an interpolation  of the specific heat data from  from the Monte Carlo curves.
We use the $C_V$ peaks to trace the phase boundaries (black points; the white dashed lines are guides to the eye) between the paramagnetic phase (PM), ferromagnetic (FM), Checkerboard (CB), and self-frustrated (SF). 
The frontier between the SF and the CB phase is an exception, with white points taken from the maximum of the order parameter fluctuations for the SF phase.
Yellow dashed lines mark the steep low temperature boundaries deduced from the measurement of entropies and energies at very low temperature. 
The vertical lines correspond to the three values of $\alpha/\alpha_c$ chosen to represent the characteristics of each phase.
Bottom panel: magnetic structure factors at different phases:
\textit{b)} ferromagnetic order (color scale 10 times bigger than the other two phases); 
\textit{c)} self-frustrated phase; 
\textit{d)} checkerboard order. 
}
\label{fig:SQ_colorCV}
\end{figure}

Our main results can be summarized by the phase diagram shown in Fig.~\ref{fig:SQ_colorCV}.  The most relevant tuning parameter in this study is $\alpha$, which controls the sensitivity of the exchange couplings to lattice distortions.  Large values of $\alpha$ correspond to materials in which small ionic displacements can induce qualitative changes in the magnetic behavior. Throughout this work, the resulting displacement values remain within the range experimentally reported for multiferroic and frustrated magnetic systems~\cite{gardner2010magnetic,slobinsky2021monopole,vignau_2023}.

We identify a critical value 
\begin{equation}
\alpha_c = \sqrt{2/5\ K_e/|J_0|}\approx 53.67 \, ,    
\label{eq:alpha_c}
\end{equation}
above which uniform ferromagnetic (FM) order is suppressed at the lowest temperatures. This value (that can be calculated following the guidelines provided in Sec.~\ref{sec:GSproperties}), together with $|J_0|$, are the natural scales that we will use for spin-phonon coupling and energy. Fig.~\ref{fig:SQ_colorCV} shows that for $\alpha/\alpha_c > 1$, the system undergoes a spontaneous distortion into a phase we term \emph{self-frustrated} (SF); it is characterized by the absence of magnetic long-range order, and will be studied in detail in Sec.~\ref{sec:SF}. At larger values of $\alpha/\alpha_c$, a second spontaneous distortion drives the system into a magnetically ordered phase with a regular up--up--down--down dimerized antiferromagnetic pattern along the three kagom\'e directions; we refer to this as the \emph{checkerboard} (CB) phase (see Fig.~\ref{fig:structures}-b).

\subsection{Specific heat and Entropy}\label{sect:CvS}

\begin{figure}[bt]
\centering
\includegraphics[width=0.95\linewidth]{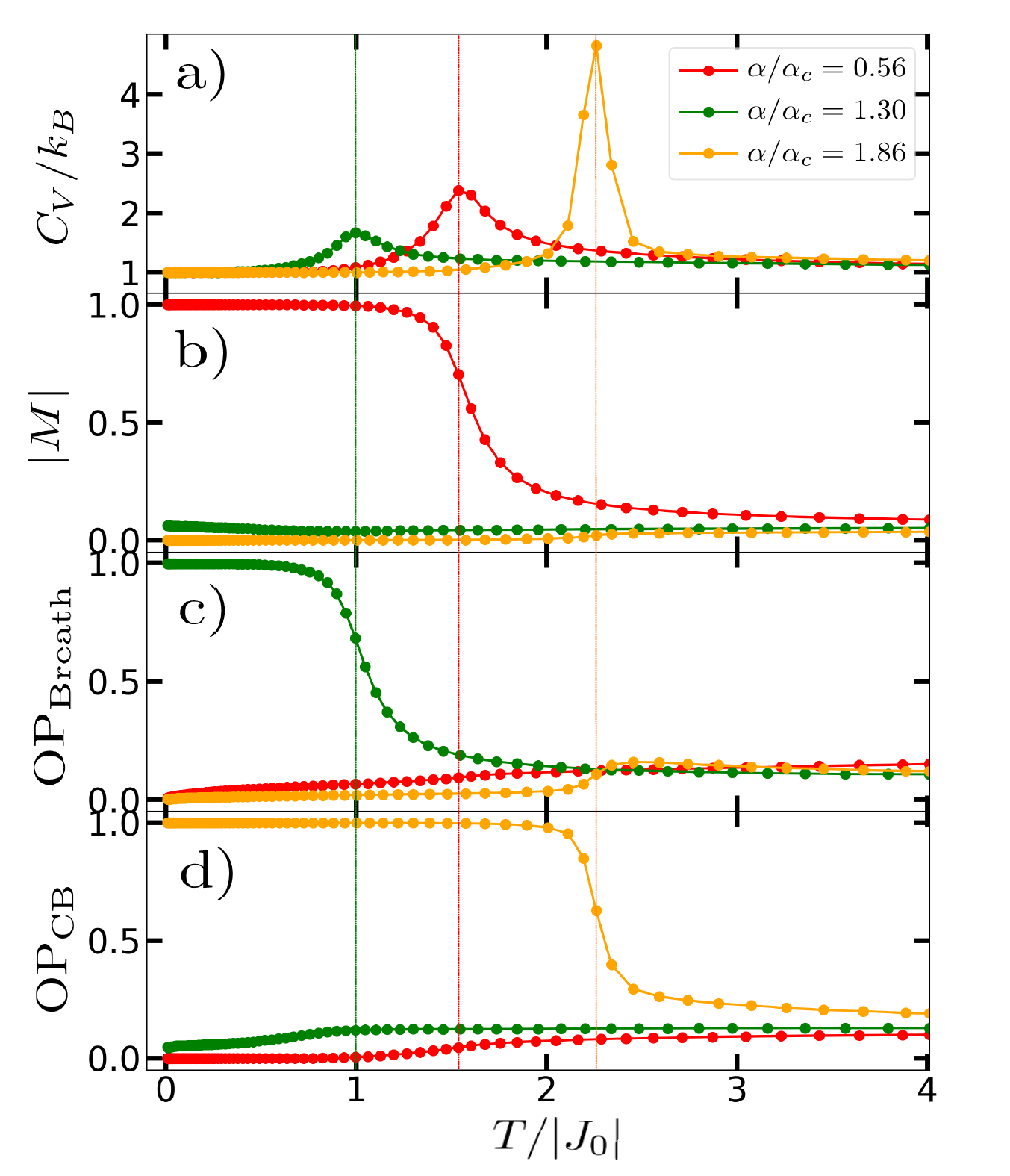}
\caption{
Panel a): 
Specific heat  vs. temperature for three representative values of the magnetoelastic coupling $\alpha/\alpha_c$ (see caption). 
The peak in $C_V$ in each curve marks a phase transition into different ordered states, characterized by suitable order parameters.
Panels b) to d):
Order parameters as a function of temperature. Each parameter characterizes a distinct type of order appearing in different ranges of the magnetoelastic coupling: the magnetization modulus $|M|$ for the FM phase at low $\alpha/\alpha_c$ (red data), the breathing parameter for the SF phase at intermediate $\alpha/\alpha_c$ (green data), and the CB order parameter for large $\alpha/\alpha_c$ (orange data).}
\label{fig:CV_S_OP}
\end{figure}

The background color map of Fig. \ref{fig:SQ_colorCV} represents an interpolation of the specific heat, computed from Monte Carlo fluctuations of both the magnetic and elastic degrees of freedom. Black dots indicate the locations of the $C_V$ peaks. Figure \ref{fig:CV_S_OP}(a) displays the specific heat as a function of temperature for three representative values of $\alpha$, marked by vertical dotted lines in the phase diagram: $\alpha/\alpha_c = 0.56,1.3,1.86$, shown in red, green, and orange, respectively. The baseline value $C_V = 1\ k_B$ (with $k_B$ the Boltzmann constant) arises from Einstein site phonons. In all cases, pronounced peaks identify the boundaries between the three distinct low-temperature phases and the high-temperature paramagnetic state. We also observe finite-size effects in each case (not shown), consistent with the occurrence of thermodynamic phase transitions.

\begin{figure}[bt]
\centering
\includegraphics[width=0.95\linewidth]{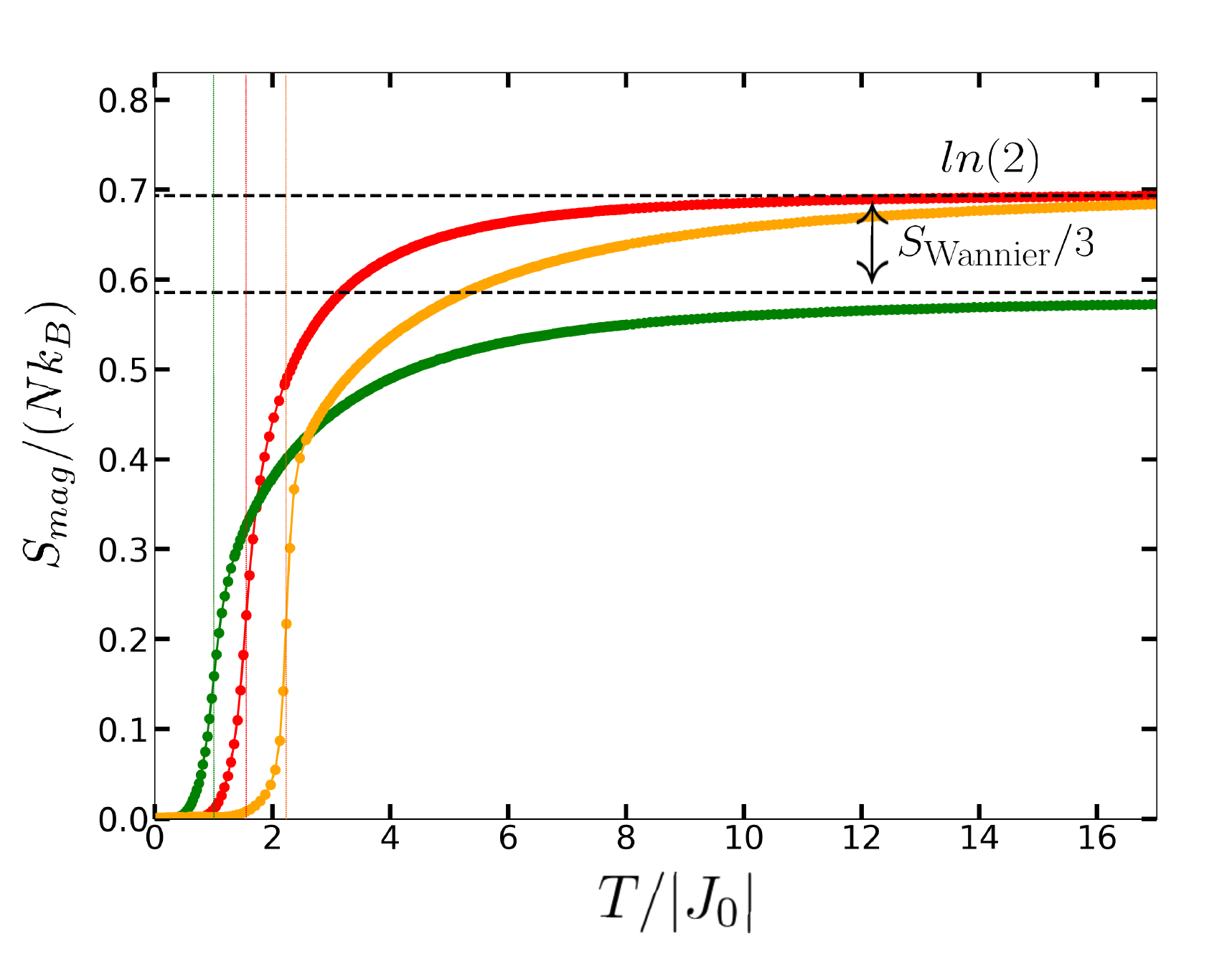}
\caption{
Entropy per spin vs. temperature for representative values of the magnetoelastic coupling $\alpha/\alpha_c = 0.56,1.3,1.86$ (using the same color code as in Fig.~\ref{fig:CV_S_OP}). The residual entropy (1/3 of the value of the residual entropy found by Wannier for the antiferromagnetic \textit{triangular} lattice) for intermediate $\alpha$ evidences that in this case the order is partial.
}
\label{fig:entropy}
\end{figure}

Further insight into the nature of these phases can be obtained by integrating the specific heat to evaluate the associated entropy change. Since we are interested in the magnetic degrees of freedom, we subtract the phonon contribution prior to performing the integration. Fig.~\ref{fig:entropy} shows the resulting entropy per spin for the three selected values of $\alpha/\alpha_c$.  For both low and high values of $\alpha/\alpha_c$, the total entropy change approaches $k_B \ln(2)$, the value expected for completely disordered Ising variables. In contrast, this is not the case for the intermediate phase, indicating a degenerate ground state. The missing entropy in this regime is very approximately one third of the Wannier residual entropy of the triangular antiferromagnetic Ising model~\cite{wannier1950antiferromagnetism}. This value suggests the emergence of an antiferromagnetic triangular structure coupling effective degrees of freedom, such as those arising from the decimation of 3-site plaquettes.

\subsection{Order parameters}

An analysis of the magnetic and elastic configurations obtained from Monte Carlo simulations motivates the introduction of three order parameters to characterize the different phases. For the ferromagnetic phase, the natural choice is the normalized modulus of the net magnetization, $|M|$, evaluated relative to its saturation value (see Fig.~\ref{fig:structures}b).

For the CB phase, we employ the order parameter $OP_{CB}$ introduced in Ref.~\cite{pili2019two}, which incorporates magnetic contributions weighted by the spatial phase factor associated with the up--up--down--down antiferromagnetic pattern of the checkerboard state (see Fig.~\ref{fig:structures}d).

Before introducing the thermodynamic order parameter for the intermediate phase, it is useful to note that the network of corner-sharing triangles forming the kagom\'e lattice can be decomposed into two triangular sublattices: one composed of ``up'' triangles (shown in green in Fig. ~\ref{fig:SF}) and the other of ``down'' triangles (shown in red or blue). The lattice distortions shown in the configuration of Fig.~\ref{fig:SF} provide then a natural criterion for defining an order parameter for this phase: the down-pointing triangles, which centers form a triangular sublattice, exhibit ferromagnetic correlations among their vertices and undergo an area contraction, while the up-pointing triangles in the complementary sublattice display the opposite behavior. Motivated by this, we introduce a \emph{breathing} order parameter, $OP_{\text{Breath}}$, defined as the normalized difference in area between the two types of triangles,

\begin{equation}
OP_{\text{Breath}}=\frac{K_e}{5\sqrt{3}\alpha|J_0|}\left\langle\left| \sum_\bigtriangleup \rm{area}(\bigtriangleup)-\sum_\bigtriangledown area(\bigtriangledown) \right|\right\rangle,  
\label{eq:OP-SF}
\end{equation} 
where the prefactor takes into account the normalization of the order parameter for different values of $\alpha$, and the brackets indicate thermal averaging. Fig.~\ref{fig:CV_S_OP} shows the temperature dependence of these order parameters for the three  selected values of the magnetoelastic coupling $\alpha/\alpha_c$.

For low spin-phonon coupling, as shown with $\alpha/\alpha_c= 0.56$ in Fig. \ref{fig:CV_S_OP}, the average magnetization $|M|$ (red points, panel b)) is saturated at low $T$ and transitions to zero (except for finite size effect) at the corresponding $C_V$ peak. Both $OP_{\text{Breath}}$ (red points, panel c) and $OP_{CB}$ (red points, panel d) vanish at low $T$.  This behavior characterizes a FM phase at low $\alpha$.

For intermediate values, here $\alpha/\alpha_c = 1.3$ (green curve), neither the magnetization $|M|$ nor the CB order parameter are significant. Instead, $OP_{\text{Breath}}$ reaches its saturation value at temperatures below the $C_V$ peak, signaling a symmetry breaking between the ``up'' and ``down'' triangle sublattices. As will be discussed afterwards, this magnetic driven transition is quite remarkable in itself in the context of the breathing kagomé systems~\cite{ezawa2018higher,bolens2019topological,wang2023quantum}.

Lastly, for large coupling values, here $\alpha/\alpha_c = 1.86$ (orange points), we only observe the CB order parameter starting its growth to saturation at temperatures below the corresponding $C_V$ peak.

We can use these order parameters to obtain a more precise determination of the boundary between the SF and CB phases. While in some cases, such as $\alpha/\alpha_c = 1.6$, two distinct peaks can be resolved in the specific heat as a function of temperature, they are generally difficult to distinguish due to finite-size rounding effects. Instead, we identify the phase boundary from the maximum in the fluctuations of the SF order parameter (white dots in Fig.~\ref{fig:SQ_colorCV}).

Finally, to determine the boundary between the FM and SF phases we equate their free energies. In order to do so, we assume that at low temperatures we can approximate both the energy and entropy by their zero-temperature values, procured as a function of $\alpha$ through the Monte Carlo simulations. The boundary obtained is indicated by a yellow line in Fig.~\ref{fig:SQ_colorCV}.

\subsection{Structure factors}

To highlight the magnetic ordering in the different low-temperature phases, we compute the static structure factor at selected values of $\alpha$. The bottom panels of Fig.~\ref{fig:SQ_colorCV} display the Fourier transform of the spin--spin correlation function at the base temperature $T/|J_0| = 10^{-2}$, for the three values of $\alpha/\alpha_c$ marked by dashed lines in the top panel, and for a system size $L = 6$. To simplify the analysis, we used the original lattice positions for the Fourier transform. The distortions modulate the peak intensities, without altering their positions.

Panels (b) and (d) show the characteristic structure-factor patterns associated with the FM and CB phases, respectively, both exhibiting long-range magnetic order. The marked difference in the height and sharpness of the peaks---note the order-of-magnitude difference in scale between panels (b) and (d)---originates from the complex $12$-spin unit cell of the CB phase.

Panel (c) shows the structure factor averaged over 200 distinct ground-state configurations in the SF phase at $\alpha/\alpha_c = 1.3$. Although the averaged structure factor preserves the sixfold symmetry of the Brillouin zone, its pattern differs markedly from those of the FM and CB phases. The resulting pattern corresponds to partial order and coincides with that of the antiferromagnetic Ising model on the triangular lattice~\cite{yoon2014diffraction}, for which the residual entropy was first calculated by Wannier. In the present case the characteristic lattice spacing is doubled, reflecting the arrangement into triangular sublattices of either the up or down pointing triangles.

\subsection{Lattice distortion patterns}

To gain further insight into the microscopic origin of the different magnetic phases, it is useful to examine in detail the ground-state distortion patterns that give rise to modified exchange couplings $J_{ij}$ according to Eq.~(\ref{eq:Jij}). To this end, we analyze the statistics of the lattice distortions obtained from Monte Carlo simulations at the lowest simulated temperature.

As expected, the low-temperature ferromagnetic phase corresponds to an undistorted system, since lattice distortions do not provide any magnetic energy gain. All exchange couplings retain the value $J_{ij}=-1$, and the magnetization remains saturated at $M=\pm1$. 

At the opposite extreme, for large values of $\alpha$, we find the regular CB distortion pattern shown in Fig.~\ref{fig:structures}(b). There, the small white arrows indicate the lattice displacements, while the plaquette colors represent the spin orientation at the corresponding vertices. In this configuration, the ferromagnetic coupling on one out of every two bonds along each of the three kagomé directions is enhanced, while the remaining bonds are sufficiently distorted to reverse their character from ferromagnetic to antiferromagnetic. The resulting network of magnetic interactions is \textit{free of frustration}, allowing the spins to arrange in an up-up-down-down sequence along each kagom\'e line. This produces a regular pattern of hexagons and triangles with opposite spin polarization.

One out of four hexagons contains short bonds associated with strong ferromagnetic couplings, thereby lowering the magnetic energy through ferromagnetic alignment (light blue plaquettes, corresponding for instance to down spins in the figure). Surrounding these hexagons are isosceles triangles with one short side belonging to the ferromagnetic hexagon and two elongated sides satisfying $\alpha \delta r_{ij}>1$, which generates antiferromagnetic couplings. The spins at the apical vertices of these triangles can then lower their energy by aligning opposite to the spins connected by the short ferromagnetic bond. These distortions, in turn, force the remaining triangles (light red in Fig.~\ref{fig:structures}(b)), which share vertices with the isosceles ones, to become equilateral with short sides. Such triangles favor fully aligned spins (up spins in the figure), oriented opposite to those on the ferromagnetic hexagons.

As mentioned, this tiling covers the kagom\'e lattice without introducing geometrical frustration. One out of every four triangles is equilateral and ferromagnetic, while the remaining three are isosceles with the geometry described above. As a consequence, the net magnetization vanishes. Crucially, the contracted ferromagnetic triangles populate in equal measure both the up down triangle sublattices. This results in a \textit{null} breathing order parameter.

We find two distinct values for the ferromagnetic couplings, corresponding to bonds on the edges of contracted hexagons and contracted triangles. On the other hand, there is a single value of antiferromagnetic couplings, associated with the elongated sides of the isosceles triangles. These couplings are accompanied by a regular bond-length dimerization along each kagom\'e direction, further modulated by a weak period-four harmonic that stabilizes the up--up--down--down magnetic order along all three kagom\'e axes. As a result, the system exhibits a unique elastic and magnetic ground state, up to $\mathbb{Z}_2$ symmetry and discrete lattice translations.

Qualitatively, at large spin--phonon coupling the system achieves a net energy gain by selectively reinforcing some ferromagnetic interactions while inducing antiferromagnetic ones, despite the associated elastic cost. We observe this low-temperature CB state over a broad range, $\alpha/\alpha_c \gtrsim 1.5$, with distortion amplitudes that increase only weakly with $\alpha/\alpha_c$.

The same CB configuration was previously reported in Ref.~\onlinecite{pili2019two} for both ferromagnetic and antiferromagnetic kagom\'e magnetoelastic systems at large spin-phonon coupling. In the antiferromagnetic case, the effective ferromagnetic couplings emerge from lattice distortions that separate spins which would otherwise interact antiferromagnetically. Once elastic distortions dominate, the sign of $J_0$ therefore becomes irrelevant.

\section{The self-frustrated phase}\label{sec:SF}

\subsection{Ground state description}

\begin{figure}[bt]
\centering
\includegraphics[width=0.8\linewidth]{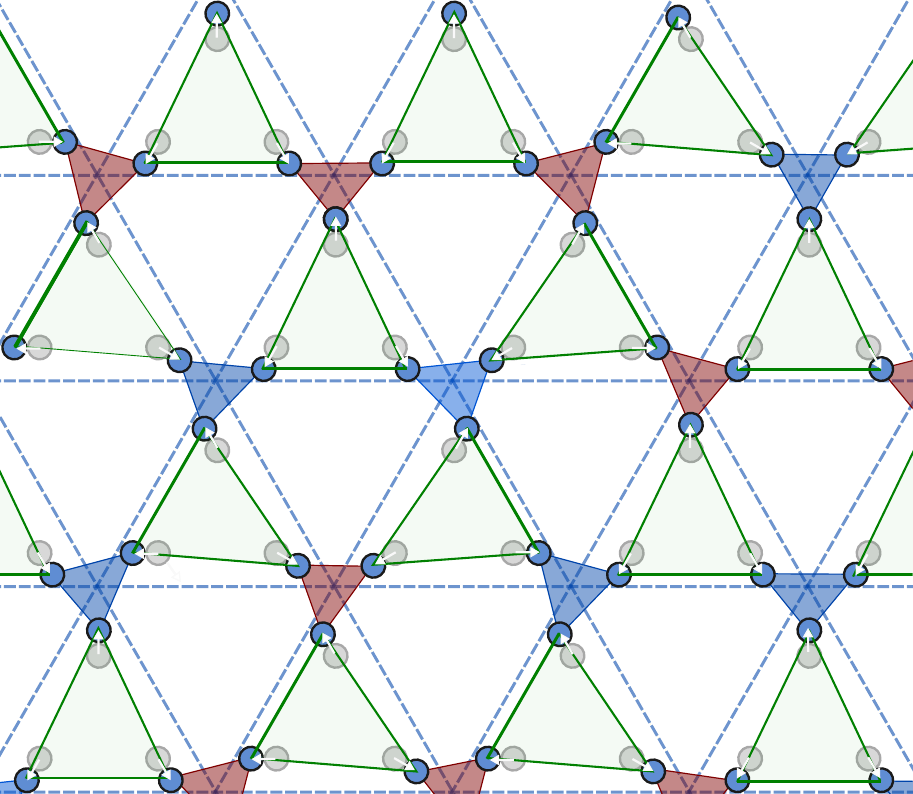}
\caption{
A microstate within the degenerate SF phase ground state. 
The spontaneous symmetry breaking is reflected in that all ``up'' triangles (green) are stretched to antiferromagnetic exchanges (AFTs), 
while all ``down'' triangles (light red and blue according to their spin orientation) are compressed reinforcing ferromagnetic couplings (FMTs). 
AFTs have isosceles distortions, with their symmetry axis lying along one of three possible directions. 
Their shape is robust: two particular values of $J_{ij} > 0$ repeat along the lattice for all AFTs (thick and thin green lines), in all microstates. 
The associated shortenings of the sides of the FMTs produce several shapes but up to five different values for the reinforced ferromagnetic couplings,
whose distribution varies among microstates.
The FMTs carry effective Ising spins $S=3$, forming a triangular lattice (light blue dashed lines) coupled antiferromagnetically.
}
\label{fig:SF}
\end{figure}

At intermediate spin--phonon coupling we observe microscopic fluctuations among a wide variety of configurations even at our lowest temperature, which is a small fraction of the characteristic magnetic energy ($T/|J_0| = 10^{-2}$). In light of the residual Wannier-like entropy discussed in Section~\ref{sect:CvS}, the number of such configurations can be expected to grow exponentially with system size. Among these, we identify microstates with different magnetizations, ranging from zero up to one third of the saturation value.  Also, as we have seen in the study of the breathing order parameter $OP_{\rm breath}$, the SF phase breaks the symmetry between the up and down triangular sublattices.

Fig.~\ref{fig:SF} illustrates a SF magnetoelastic configuration in which the ``down'' triangles {\em contract} into a variety of shapes. We term them \textit{ferromagnetic triangles} (FMTs), since they host ferromagnetically aligned spins at their three vertices; their spin orientation can be either up (red in the figure) or down (blue).  Conversely, the ``up'' triangles in this configuration {\em expand} into the isosceles shape: the two equal sides become longer than the third (thick line in the figure), yet the distortion is sufficient to induce antiferromagnetic bonds along all three edges. We refer to these as \textit{antiferromagnetic triangles} (AFTs).
Because FMTs and AFTs always share vertices, the resulting structure exhibits a nonuniform bond dimerization along each kagom\'e line, namely an alternation of short (ferromagnetic) and long (antiferromagnetic) bonds.

As it is quite well known, we cannot expect the spins at the vertices of the AFTs to simultaneously satisfy all three antiferromagnetic couplings; this gives rise to intrinsic self-frustration. In this case, due to the different $J_{ij}$ values, frustration is partial. Spins connected by the two longer, symmetric bonds tend to align antiparallel, at the expense of accommodating a parallel alignment along the weaker antiferromagnetic side (thick lines in Fig.~\ref{fig:SF}). 
In these configurations, the FMTs carry a net magnetization of \(\pm 3\) (in spin units), while the AFTs carry \(\pm 1\). In contrast to the CB phase, the mere alternation of ferromagnetic and antiferromagnetic bonds does not suffice to stabilize an up--up--down--down magnetic order.

Geometrically, the AFTs are robust: they all adopt the same isosceles shape and size, giving rise to only two distinct antiferromagnetic couplings. While the orientation of their symmetry axes may vary, the values of these couplings remain identical across all AFTs and for all microstates at a given $\alpha/\alpha_c$. In contrast, the ionic displacements at the vertices of the FMTs are determined by the arrangement of the surrounding AFTs, leading to a variety of shapes. In particular, between zero and three apical vertices of neighboring isosceles triangles may converge on a given FMT. As a result, one finds up to five distinct reinforced ferromagnetic couplings (i.e., five different short-bond lengths); the distribution frequency of these five values varies from one microstate to another.  

\begin{figure}[bt]
\centering
\includegraphics[width=0.95\linewidth]{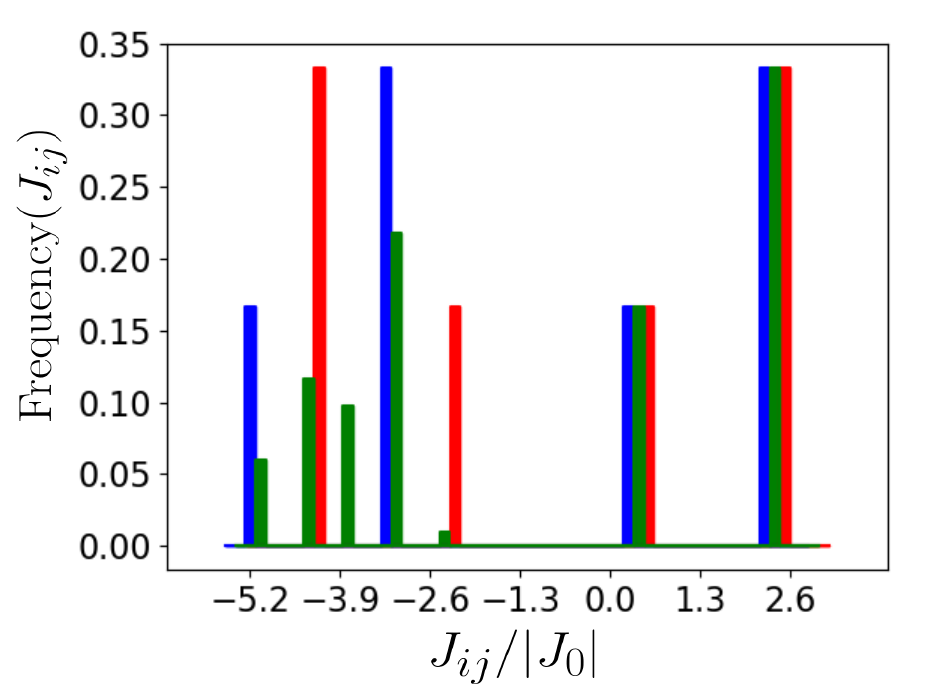}
\caption{
Histograms for the magnetic exchange couplings distribution measured for three different low energy configurations (in different colors and slightly displaced for clarity)  of the SF phase for $\alpha/\alpha_c=1.3$ at $T=10^{-4}$. 
Bar heights are normalized such that frequencies add to one for each microstate. In spite of the disorder and the continuous nature of the variable, the AF bonds (with $J_{ij}>0$) are restricted to only two values with rational frequencies $1/3$ and $1/6$. 
This restriction operates not only within a given configuration, but for the whole macrostate, and is related to the robustness of the isosceles AFTs. 
On the other hand the ferromagnetic $J_{ij}>0$ can take five different values with continuous frequencies. Within the disordered macrostate, there exist regularly ordered configurations (two of which are represented in blue and in red here) with less than 5 values for $J_{ij}<0$.
}
\label{fig:SF-exchanges}
\end{figure}

Fig.~\ref{fig:SF-exchanges} shows the histograms of the magnetic exchange couplings measured for three representative SF configurations, each represented by a different color. The identical length and position of color bars on the right (the slight displacement along the $x-$axis for the different configurations was introduced for visual clarity) illustrates the aforementioned robustness of the AF couplings. On the other hand, although the five negative exchange values are the same for every configuration, the distribution in ferromagnetic couplings is observed to be quite different for each microstate. There are two different pairs of values for the blue and red case, and all five ferromagnetic $J_{ij}$ for the green one. 

From this description, one can identify a global \(\mathbb{Z}_2\) variable that selects whether the ``up'' or ``down'' triangles act as FMTs, as well as a local effective Ising variable associated with each FMT. The peak observed in \(C_V\) versus \(T\) for $1 < \alpha/\alpha_c < 1.5$ signals the spontaneous breaking of this global $\mathbb{Z}_2$ symmetry, which is also reflected in the behavior of the breathing order parameter.

Effectively, the FMTs can be viewed as composite moments in an effective triangular lattice with antiferromagnetic interactions. This description captures the frustrated character of the state and accounts for the emergence of a residual entropy equal to one third of the Wannier value for the triangular antiferromagnet~\cite{wannier1950antiferromagnetism}. This interpretation is further supported by the structure factor shown in Fig.~\ref{fig:SQ_colorCV}c), which captures the spin correlations of the effective moments on the triangular lattice. 

\subsection{Ground state properties: robust antiferromagnetic triangles and degeneracy}\label{sec:GSproperties}

The robustness of the AFTs is a central feature of the SF phase at low temperatures: their shape at low $T$ remains unique within thermal resolution. Up to the breaking of reflection symmetry, the different microstates correspond to distinct ways of assembling the AFTs in one triangular sublattice of the kagomé into configurations that determine the disordered FMTs in the other sublattice: the same rigid building blocks can be combined into different ``puzzles'', all with the same energy. This leads to an exponentially large number of configurations in which the FMT couplings can take up to five distinct values, appearing with different frequencies across microstates (Fig.~\ref{fig:SF-exchanges}). 

The rigidity of the AFTs may appear counterintuitive in view of the structural and magnetic disorder present in the SF phase. Moreover, the nature of the  degeneracy of its ground-state manifold calls for a deeper analysis. To address these issues, we resort to analytical methods together with a modified Monte Carlo approach.

\subsubsection{Analytical approach: vector linear approximation for  $J_{ij}$ couplings} \label{sec:foApp}

The spin phonon coupling term 
$\alpha (r_{ij}-1) S_i S_j$ 
in Eq.~(\ref{eq:H})  
arises from the linear expansion of the exchange couplings 
$J_{ij}(\mathbf{r}_{ij})$ 
in  Eq.~(\ref{eq:Jij}) for small distance distortions $(r_{ij}-1)$, assuming isotropy. 
Further analytical discussion can be done by expanding  
\begin{equation}
    r_{ij} \approx 1+\hat{e}_{ij} \cdot (\delta \mathbf{r}_j-\delta \mathbf{r}_i)  \, ,
\end{equation}
where \(\hat{e}_{ij}\) is the unit vector pointing from site \(i\) to site \(j\) in the pristine lattice, and terms of order \((\delta \mathbf{r}_i)^2\) and higher are neglected. Appendix \ref{Ap:LinAp} details the derivation and shows that equilibrium solutions are obtained with $|\delta \mathbf{r}_i| \propto \alpha/K_e$. The use of the vector linear approximation is therefore well justified in this context. 
For our largest values of $\alpha \sim 10^{2}$ and $K_e = 7200$, one obtains $|\delta \mathbf{r}|$ of order $10^{-2}$ in units of the lattice parameter. This lies within the range of displacements observed in multiferroics and certain frustrated systems~\cite{gardner2010magnetic,slobinsky2021monopole,vignau_2023}.

In Appendix~\ref{Ap:LinAp} we derive, within the present approximation, the force exerted by a neighboring spin $j$ on a given spin $i$, assuming the spins $S_i$ and $S_j$ are known. It can be written (in units of $|J_0|$) as
\begin{equation}
    \mathbf{f}_{ji} = -{\rm sgn}(J_0)\,\alpha \, S_i S_j \, \hat{e}_{ij} \,.
\end{equation}
This force is directed along the line connecting the spin sites \textit{in the undistorted lattice}. As expected, in the ferromagnetic case ($\mathrm{sgn}(J_0) = -1$) nearest-neighbor ions experience an attractive (repulsive) force when their spins are parallel (antiparallel).

The equilibrium configuration (i.e., the minimum of the total energy at $T=0$) is obtained by finding the displacement $\delta \mathbf{r}_i^{\mathrm{eq}}$ at which the isotropic elastic force balances the magnetic force $\bm{\mathcal{F}}_i \equiv \sum_j \mathbf{f}_{ji}$ exerted by the four spins $j$ neighboring site $i$:
\begin{equation}
    \delta \mathbf{r}_i^{eq} = \frac{\bm{\mathcal{F}}_i }{K_e} \, .
\end{equation}

\begin{figure}[bt]
\centering
\includegraphics[width=0.95\linewidth]{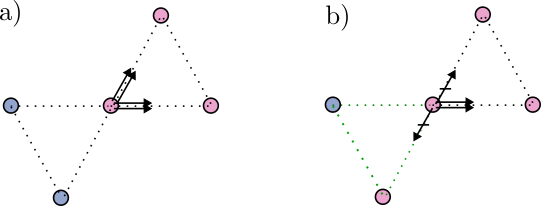}
\caption{
Forces exerted on a given magnetic ion by its neighbors, in the  vector linear approximation. 
Any site is a shared corner of a FMT (upper right in the figures) and an AFT (lower left).
Spin orientation is given by the circle colors (say up in light red, down in light blue).
One of the shown possibilities describes any site in any of the SF microstates.
a) the shared corner is the apical vertex of the AFT: the net force points to the center of the FMT.
b) the shared corner is a basal vertex of the AFT: the net force points along one of the sides of the FMT.
}
\label{fig:forces}
\end{figure}
In the SF configurations, any given spin $S_i$ is shared by an FMT and an AFT; its orientation is obviously dictated by the FMT. Two relevant situations arise, depending on the orientation of the other two spins in the AFT:

\par \textit{a}) If $S_i$ sits at the apical vertex of an AFT (see Fig.~\ref{fig:forces}a)), the spins at the basal vertices are opposite. In this case, the neighboring FMT spins attract $S_i$, while the AFT spins repel it with equal magnitude. The resulting net force points toward the center of the (pristine) FMT, producing a displacement of magnitude $\delta r_i^{\mathrm{eq}} = 2\sqrt{3}\alpha/K_e$.

\par \textit{b}) If $S_i$ sits at a basal vertex of an AFT (see Fig.~\ref{fig:forces}b)) the other basal spin has the same orientation, while the apical one is opposite. In this situation, the attraction due to the second AFT basal spin cancels that of the FMT spin along the same kagom\'e line, whereas the repulsion from the AFT apical spin adds to the attraction from the remaining FMT spin. The net force points away from the apical AFT vertex, along the corresponding kagom\'e line, resulting in a displacement $\delta r_i^{\mathrm{eq}} = 2\alpha/K_e$.

The key point underlying the robustness of the AFTs and the associated antiferromagnetic $J_{ij}$ in the SF phase is that any spin can be regarded as a vertex of an AFT and therefore falls into one of the two situations described above, regardless of the specific SF configuration. Within the vector linear approximation, the displacements of both basal and apical vertices are thus completely determined and do not depend on how the AFT ``puzzle'' is assembled.

Once the ionic displacements $\delta r_i^{\mathrm{eq}}$ are determined, the zero-point energy $E(T=0)$ can be computed for any SF microstate. Appendix~\ref{Ap:LinAp} presents a detailed derivation, leading to the expression
\begin{equation}
    \frac{E(T=0)}{|J_0|} \approx {\rm sgn}(J_0) \sum_{\langle i,j \rangle} S_i S_j - \sum_i \frac{K_e}{2} (\delta r_i^{eq})^2 \, .
    \label{eq:E_T0_fo}
\end{equation}
In the first term, the summation can be carried out straightforwardly, since in any configuration half of the triangular plaquettes are FMTs and the other half are AFTs (see Fig.~\ref{fig:SF}). The second summation can likewise be evaluated independently of the specific configuration: every spin occupies either an apical or a basal vertex of an isosceles AFT, for which the corresponding $|\delta r_i^{\mathrm{eq}}|$ has already been determined. Two thirds of the sites correspond to basal AFT vertices, while the remaining third are apical vertices. As a result, within the vector linear approximation, the magnetoelastic energy is identical for all microstates within the SF ground-state manifold.

Eq.~(\ref{eq:E_T0_fo}) allows also for the calculation of the magnetoelastic energy (within the linear approximation) for any microstate at $T=0$; in particular, for the ordered configurations associated with the other two phases. By equating these energies we can find both transition points at $T=0$ and thus the expression for $\alpha_c$ given in Eq.~(\ref{eq:alpha_c}).

\subsubsection{Monte Carlo simulation of distortions in the SF phase}

In the previous section, terms of order $\left(\delta r_i^{\mathrm{eq}}\right)^2$ were neglected. As noted above, we estimate the dimensionless displacements to be $|\delta r_i^{\mathrm{eq}}| \approx 10^{-2}$ at the largest values of $\alpha$ considered in this work. Accordingly, one expects energy corrections of order $E/|J_0| \sim 10^{-4}$. Resolving such second-order effects in Monte Carlo simulations would therefore require temperatures at least an order of magnitude lower than our minimum value, $T/|J_0| = 10^{-2}$.

Such low temperatures can be accessed within our available computational resources by fixing the spin configuration and simulating only the elastic degrees of freedom. For various values of the magnetoelastic coupling $\alpha/\alpha_c$ within the SF regime, we adopt the following strategy: we select several fixed spin configurations (i.e., different solutions of the AFT ``puzzle'') and perform Monte Carlo simulations of the full Hamiltonian in Eq.~(\ref{eq:H}) to determine the equilibrium displacements, at temperatures as low as $T/|J_0| = 10^{-4}$.
 
With this procedure, we find that the combined magnetic and elastic energies of these configurations are essentially identical, with only small residual differences. While we cannot determine the exact ground-state degeneracy of the SF phase, we can place bounds on its energy split. The lowest-energy configuration found corresponds to a regular arrangement of the AFT isosceles triangles in which FMTs formed by apical spins are surrounded, on their triangular lattice, exclusively by FMTs composed of basal spins. In this configuration, both types of FMTs remain equilateral, albeit with distinct degrees of contraction. The highest energy in the manifold, by contrast, is found for another regular configuration in which all AFT symmetry axes are aligned, leading to contracted isosceles FMTs. Nevertheless, the energy difference between these extreme cases remains very small; for a representative value $\alpha = 1.3\,\alpha_c$, we find $\Delta E/|J_0| \approx 10^{-4}$. This explains why entropy estimates obtained from specific-heat calculations—performed with both magnetic and elastic updates and reaching temperatures down to $T/|J_0| \approx 10^{-2}$ give the macroscopic residual entropy of the effective triangular antiferromagnetic lattice.

\section{Summary, conclusions, and perspectives}

We have shown that magnetoelastic coupling can dynamically generate magnetic frustration in an otherwise unfrustrated system, leading to a partially ordered phase with apparent residual entropy. In contrast to the conventional scenario, where lattice distortions relieve frustration by selecting an ordered state, here the system lowers its energy by developing distortions that induce competing interactions and an emergent frustrated state.

Specifically, we studied a ferromagnetic Ising model on the kagom\'e lattice coupled to elastic degrees of freedom. As the magnetoelastic coupling increases, the system undergoes a structural transition from a uniform ferromagnetic phase (FM) to an intermediate self-frustrated (SF) regime, before reaching a fully ordered checkerboard (CB) phase at stronger coupling. The SF phase consists of a disordered arrangement of ferromagnetic triangles that act as effective composite spins on an antiferromagnetic triangular lattice. This effective description explains the absence of long-range order, the algebraic correlations observed in Fig.~\ref{fig:SQ_colorCV}c), and the finite residual entropy proportional to the Wannier value. At the microscopic level, the self-frustration originates from the spontaneous formation of robust isosceles distortions, which generate frustrated antiferromagnetic interactions on a subset of bonds and thereby driving the system to self-organize into an effective frustrated magnetic model.

Within a linear approximation, we have shown that the magnetoelastic energy is degenerate over a large manifold of configurations, providing a natural explanation for the observed residual entropy and partial order. A more refined analysis indicates that nonlinear terms may, at least partially, lift this degeneracy. From our estimates, the associated energy scale is bounded by $E/|J_0| \approx 10^{-4}$. Following Ramirez~\cite{ramirez1994strongly}, one may define a frustration parameter $f$ for the SF phase as the ratio between the characteristic interaction energy and the ordering scale, yielding $f \gtrsim 10^{4}$. This is indeed quite large if we consider that values above $10$ are considered strongly frustrated~\cite{ramirez1994strongly}.

Our results illustrate a general mechanism by which frustration can emerge dynamically from the interplay between magnetic and elastic degrees of freedom. This raises the possibility that similar self-generated frustration may occur in other magnetoelastic systems, even when the underlying magnetic interactions are not intrinsically competing.

Recent years have witnessed a growing interest in kagomé-based materials exhibiting structural distortions that break the equivalence between up and down triangles, commonly referred to as breathing kagomé lattices~\cite{ezawa2018higher,bolens2019topological,wang2023quantum,ogunbunmi2023magnetic}. 
In these systems, inequivalent exchange couplings on small and large triangles, typically associated with trimerization, lead to a rich interplay between lattice, spin, and electronic degrees of freedom, often accompanied by magnetoelectric effects and, in some cases, multiferroicity~\cite{li2021two}. 
Representative examples include van der Waals and Janus kagomé materials such as Nb$_3$X$_8$ compounds and related systems, where out-of-plane polarization and electric-field control of the breathing distortion have been reported~\cite{wang2026breathing,xie2025manipulation}. 
Within this context, the SF phase uncovers a qualitatively similar scenario: the system develops an emergent --breathing-- asymmetry between the two triangle sublattices but without forming a regular pattern. 
While one set of triangles effectively contracts into ferromagnetic units and the complementary set expands and hosts frustrated interactions, 
the phenomenon leads to a partially ordered state. 
Therefore, while conventional breathing kagomé systems generally rely on preexisting structural inequivalence between triangular units~\cite{ezawa2018higher, bolens2019topological}, the intermediate phase reported here highlights a distinct mechanism in which such differentiation emerges dynamically and is intrinsically tied to the generation of frustration and, potentially, to magnetoelectric responses.

The spontaneous differentiation between inequivalent structural units induced by magnetoelastic coupling bears some conceptual resemblance to bond-disproportionated phases in rare-earth nickelates \cite{zhou2004chemical,subedi2015low,green2016bond}, where lattice distortions generate inequivalent octahedra coupled to electronic degrees of freedom. However, in contrast to the long-range ordered breathing pattern observed in nickelates, closer in spirit to our checkerboard phase, the intermediate self-frustrated phase reported here remains disordered and is intrinsically tied to the emergence of magnetic frustration and residual entropy.

Several open questions remain. It would be interesting to explore the robustness of the self-frustrated phase beyond the present model assumptions, including different forms of spin-lattice coupling or lattice dynamics. The effect of external perturbations, such as magnetic fields, may also provide further insight into the nature of the partially ordered state. Finally, connections with other systems exhibiting emergent frustration~\cite{strevcka2012spin,strevcka2019anomalous} or residual entropy deserve further investigation.

\begin{acknowledgments}

We acknowledge the Agencia Nacional de Promoción Científica y Tecnológica (ANPCyT) Argentina, for awarding grants  PICT 2022-11-00046 and PICT 2022-11-00100, which, although financed by the Inter-American Development Bank (IDB), were later discontinued following changes in national funding policy. The authors are partially supported by Consejo Nacional de Investigaciones Cient\'ificas y T\'ecnicas (CONICET), Argentina. D.C.C. is grateful to F. Mil\'a for valuable discussions.
\end{acknowledgments}

\appendix
\section{Vector linear approximation to the Hamiltonian}\label{Ap:LinAp}

We start from the Hamiltonian in Eq.~(\ref{eq:H}). Given the dependence of $J_{ij}$ on $r_{ij}$ shown in Eq.~(\ref{eq:Jij}), 
we expand the distances $r_{ij}$ to first order in $\delta r_{i,j}$. 
Nearest neighbors in the pristine lattice are separated by unitary vectors 
$\hat{e}_{ij}=\mathbf{r}^0_{j}-\mathbf{r}^0_{i}$ 
(see Fig.~\ref{fig:structures}a)). 
A linear expansion of the distance $r_{ij}$ on vector distortions $\delta \mathbf{r}_i$, $\delta \mathbf{r}_j$ reads
\begin{equation}
    r_{ij} \approx 1+\hat{e}_{ij} \cdot (\delta \mathbf{r}_j-\delta \mathbf{r}_i)  \, .
\end{equation}
Using this approximation in the magnetic part of the dimensionless Hamiltonian in Eq.~(\ref{eq:Jij}) we obtain
\begin{align}
     &  \frac{\mathcal{H}_{mag}}{|J_0|} \equiv  \sum_{\langle i,j \rangle} J(r_{ij}) S_i S_j \nonumber \\
     & \approx {\rm sgn}(J_0) \left( \sum_{\langle i,j \rangle} S_i S_j - \alpha \sum_{\langle i,j \rangle}  \hat{e}_{ij} \cdot \left( \delta \mathbf{r}_j-\delta \mathbf{r}_i\right )) S_i S_j \right ).    
\end{align}
The first term is just the exchange energy in the pristine, undistorted lattice. 
The second one can be reordered as a sum over sites $i$, each one  interacting with its four nearest neighbors $N\!N(i)$.
We arrive to 
\begin{equation}
    {\rm sgn}(J_0) \left [\sum_{\langle i,j \rangle} S_i S_j +  \alpha \sum_i \left(\sum_{j \in N\!N(i)} S_i S_j \hat{e}_{ij}\right) \cdot \delta \mathbf{r}_i \right]\,
    \label{eq:App_H_fo}
\end{equation}
where it is easier to identify  linear terms in vector displacements $\delta \mathbf{r}_i$.

If we assume a \textit{magnetic} configuration $\{S_i\}$ it is straightforward to deduce 
the displacements $\{\delta \mathbf{r}^{eq}_i\}$ that minimize the total energy. 
The first term in these equations is just a constant, 
the second one contains the  forces $\mathbf{f}_{ji}$ exerted by nearest neighbor sites $j$ on site $i$,
\begin{equation}
    \mathbf{f}_{ji} = -{\rm sgn}(J_0)\alpha S_i S_j \, \hat{e}_{ij} \,.
\end{equation}

The Hamiltonian to first order in vector distortions can also be written in terms of these forces as
\begin{equation}
    \frac{\mathcal{H}_{mag}}{|J_0|}\approx {\rm sgn}(J_0)\sum_{\langle i,j \rangle} S_i S_j - \sum_i \bm{\mathcal{F}}_i \cdot \delta \mathbf{r}_i \,
    \label{eq:App_H_fo_forces}
\end{equation}
where 
$
    \bm{\mathcal{F}}_i= \sum_{j\in NN(i)} \mathbf{f}_{ji}
$   
is the net force exerted on site $i$.

\subsubsection{Equilibrium displacements $\delta \mathbf{r}_i^{eq}$}

The structural equilibrium point for a given spin configuration can be found by completing squares in Eq.~(\ref{eq:H}) 
in order to get the total (magnetic plus elastic) energy minimum at $T=0$.
Alternatively, we found it instructive to write the equilibrium displacements $\delta \mathbf{r}_i$ at which the local elastic force  cancels out the magnetic force $\bm{\mathcal{F}}_i$,
\begin{equation}
    \delta \mathbf{r}_i^{eq} = \frac{\bf{\mathcal{F}}_i}{K_e} \, .
    \label{eq:Ap_req}
\end{equation}
Since the elastic force in our model is isotropic, the displacement direction is given by the magnetic force $\bm{\mathcal{F}}_i$. 
We stress here that $\delta r_i^{eq}$ is proportional to $\alpha/K_e$, 
thus the order of magnitude of lattice distortions can be directly estimated from the model parameters. The vector linear approximation can be seen as a first order expansion in $\alpha/K_e$.

\subsubsection{Zero temperature total energy for a given magnetic configuration}

Introducing the equilibrium values of displacement at $T=0$ (Eq.~(\ref{eq:Ap_req})) into Eq.~(\ref{eq:App_H_fo}) we can now calculate the total energy  $E(T=0)$ within the linear vector approximation,
\begin{equation}
    \frac{E(T=0)}{J_0} \approx {\rm sgn}(J_0) \sum_{\langle i,j \rangle} S_i S_j - \sum_i \frac{K_e}{2} (\delta r_i^{eq})^2 \,.
    \label{eq:App_E_T0_fo}
\end{equation}
This expression contains a  magnetoelastic correction of order $\alpha^2/Ke$ to the magnetic energy of the undistorted lattice.

\bibliographystyle{apsrev4-2}
\bibliography{postdoc}

\end{document}